\documentclass[a4paper,fleqn,usenatbib]{mnras}

\usepackage{newtxtext,newtxmath}

\usepackage[T1]{fontenc}
\usepackage{ae,aecompl}

\usepackage{graphicx}	
\usepackage{amsmath}	
\usepackage{amssymb}	
\usepackage{tablefootnote}
\usepackage{lscape}
\usepackage{longtable}
\usepackage{natbib}
\usepackage{color}
\usepackage{array} 

\newcommand{\cafex}{CAFE\textit{xtractor}}
\newcommand{\cafet}{CAFE$_2$}

\title[CAFE$_2$: the new capabilities of the CAFE spectrograph]{CAFE$_2$: an upgrade to the CAFE high-resolution spectrograph. Commissioning results and new public pipeline}

\author[Lillo-Box et al.]{
J.~Lillo-Box,$^{1,2}$\thanks{E-mail: jlillo@cab.inta-csic.es}
J.~Aceituno,$^{3}$ 
S.~Pedraz,$^{3}$
G.~Bergond,$^{3}$
D.~Galad\'{\i}-Enr\'{\i}quez,$^{3}$
\newauthor{M.~Azzaro,$^{3}$
B.~Arroyo-Torres,$^{3}$
A.~Fern\'andez-Mart\'in,$^{3}$
A.~Guijarro,$^{3}$
R.~P.~Hedrosa,$^{3}$}
\newauthor{I.~Hermelo,$^{3}$  
F.~Hoyo,$^{3}$
P.~Mart\'in-Fern\'andez,$^{3}$}
\\
$^{1}$Depto. de Astrof\'isica, Centro de Astrobiolog\'ia (CSIC-INTA), ESAC campus 28692 Villanueva de la Ca\~nada (Madrid), Spain \\
$^{2}$European Southern Observatory, Alonso de Cordova 3107, Vitacura Casilla 19001, Santiago 19, Chile \\
$^{3}$Centro Astron\'omico Hispano en Andaluc\'\i a, Observatorio de Calar Alto, Sierra de los Filabres s/n, 04550 G\'ergal (Almer\'\i a)
}

\date{Accepted for publication in MNRAS on 19-11-2019}

\pubyear{2019}

\begin{document}
\label{firstpage}
\pagerange{\pageref{firstpage}--\pageref{lastpage}}
\maketitle

\begin{abstract}
CAFE is a high-resolution spectrograph with high-precision radial velocity capabilities mounted at the 2.2m telescope of Calar Alto Observatory. It suffered from strong degradation after 4 years of operations and it has now been upgraded. The upgrades of the instrument (now named CAFE$_2$) aimed at {recovering} the throughput and {improving} the stability thanks to the {installation} of a new grating, an active temperature control in the isolated coud\'e room, and a new scrambling system. In this paper, we present the results of the new commissioning of the instrument and a new pipeline (CAFExtractor) that provides the user with fully reduced data including radial velocity measurements of FGK dwarf stars. The commissioning results show a clear improvement in the instrument performance. The room temperature is now stabilized down to 5 mK during one night and below 50 mK over two months. CAFE$_2$ now provides 3 m/s precision on the reference ThAr frames and the on-sky tests provide a radial velocity precision of 8 m/s during one night (for $S/N>50$). The throughput of the instrument is now back to nominal values with an efficiency of around 15\% at 550 nm. The limiting magnitude of the instrument for a 1h exposure and S/N=20 is V=15. With all these properties, CAFE enters into the small family of high-resolution spectrographs at 2-4m telescopes capable of reaching radial velocity precisions below 10 m/s.

\end{abstract}

\begin{keywords}
Instrumentation: spectrographs -- Techniques: radial velocities, spectroscopic -- Planets and satellites: detection
\end{keywords}



\section{Introduction}

High-resolution spectrographs with resolving power in the range $R=60000-80000$ have been extensively used in the past years for the detection and characterization of new extrasolar planets through the radial velocity technique \citep[e.g., ][]{courcol15, santerne16b,  lillo-box16c} and the vetting of planet candidates \citep[e.g., ][]{santerne15} detected by space-based photometric missions like Kepler \citep{borucki10} or TESS \citep{ricker14}. Additionally, many other science cases can take profit of this resolution, like the study of element abundances in young stars \citep[e.g., ][]{huelamo18}, the study of close binaries \citep[e.g., ][]{lillo-box15a}, open clusters \citep[e.g., ][]{casamiquela16}, massive stars \citep[e.g., ][]{holgado18} or the detection of central stars in planetary nebula \citep[e.g., ][]{aller18}. Also, attached to 2-4\,m class telescopes, these instruments can be used for extragalactic studies. 

The CAFE spectrograph \citep{aceituno13} was installed in the 2.2m telescope of the Calar Alto Observatory (CAHA, Almer\'ia, Spain) in May 2012. The instrument is fed by the telescope through a circular 1.5 arcsec diameter fiber entrance which transports the light through 18 meters from the Cassegrain focus to the coud\'e room. The room is located below the main telescope structure, separated from the rest of the building providing it with isolation of any mechanical vibration. Also, a sophisticated pneumatic stabilization system is installed below the optical bench. This guarantees a much better stability of the instrument, increasing its performance and accuracy for velocity measurements of fainter objects. CAFE had no movable pieces and a fixed spectral coverage in the range $\sim$4000-9500 \AA, with a spectral resolution of $R = 62\,000\pm5\,000$. The chamber was monitored in temperature, pressure, and humidity to check for possible relevant changes during the night. These properties allowed the instrument to reach the 20 m/s precision in radial velocity after some inter-night corrections mainly due to differences in the signal-to-noise ratio (S/N) of the different spectra of the same object. As an example, the reflex motion of the planet Kepler-91\,b \citep{lillo-box14} on its giant star was detected in \cite{lillo-box14c} with an amplitude of 60 m/s, which allowed the characterization of its orbital properties and to confirm the previously determined mass using ellipsoidal modulations. The instrument covered a substantial fraction of the telescope time at the 2.2m telescope (35\%-60\%, depending on the semester) and, since the beginning of operations of the instrument, it has been the most demanded one at the 2.2m CAHA telescope.

However, soon after the first year of operations, the efficiency of the instrument started to drop, decreasing in 2015 down to {one third} of the original values. After an intervention, the reason of this flux losses pointed to the clear degradation of the echelle grating, probably due to some sulfurous components that could be present in the air inside the chamber. As a consequence a degradation of the silver coating on the grating (maybe also aging) is the most plausible explanation.

The solution to this problem came with a new financial support by FEDER/MINECO funds, which allowed the purchase of a new grating as well as other additional improvements to the instrument such as an active temperature control of the isolated chamber. In this paper, we describe the different improvements installed and their performance over the commissioning period. Additionally, we have developed a new set of automatic scripts to monitor the health of the instrument\footnote{\url{https://github.com/jlillo/CAFE_HealthChecks}} and a new public pipeline to fully reduce the data. In section 2, we describe the main technical improvements of \cafet\ and its new characteristics. In section 3, we describe the new developed pipeline (\cafex) by outlining the basic steps; in section 4 we focus on the scientific performance of the instrument after the upgrade by showing the results of the commissioning data.

\section{New characteristics of  CAFE$_2$}

\subsection{The new grating and wavelength coverage}

\begin{figure}
\centering
\includegraphics[width=0.45\textwidth]{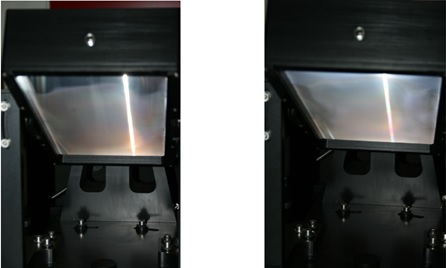}
\caption{View of the previous CAFE grating in two different dates, {Jan 2014 on the left panel and Dec 2014 on the right panel. In the left panel, it is clearly visible how the degradation of the coating already affected most of the reflective surface of the grating. After 12 months, the degradation covers the whole grating. }}
\label{fig:grating}
\end{figure}

A sensitivity problem was detected in CAFE data by early 2014, after almost two years of operations. Some sulfurous components could be present in the air because some silicone residues were found in the inner part of the protected cabin of the instrument. {Added to this, the grating was installed facing down, hence standing on its front borders. This constant force might have cracked the thin film covering the silver, thus exposing it to the open air possibly causing its degradation.} The effect is shown in Figure~\ref{fig:grating}. 
 
The degradation of the grating coating produced a drop in S/N of a factor of 3, which jeopardized the operations of the instrument. Although the degradation was not expected to continue, action to replace the grating was necessary. In May 2018, a major intervention
was performed on the instrument, which included the substitution of the grating and the removal of the silicone residues in the inner cabin of the spectrograph. The features of the new grating were selected to be the same as the original piece (i.e Newport corporation manufacturer, engraved 31.6g/mm and a Blaze angle of 63.9$^{\circ}$, see Table~\ref{tab:CAFEgrating}). The new grating was successfully installed on May 2018 and supplies a similar performance in terms of sensitivity and resolution to the original configuration.

\begin{table}
\setlength{\extrarowheight}{1pt}
\caption{Characteristics of the new grating of CAFE}              
\label{tab:CAFEgrating}      
\centering                                      
\begin{tabular}{l  l}          
\hline                        
Manufacturer	& NewPort \\
Groove density	& 31.6\,g/mm\\
Blaze angle	& 63.9$^{\circ}$\\
Substrate dimension	& 165$\times$320$\times$50\,mm\\
Clear Aperture	& 154$\times$206\,mm\\
Material	& Zerodur\\
Coating	& Protected silver\\

\hline                                             
\end{tabular}
\end{table}

\begin{figure}
\centering
\includegraphics[width=0.5\textwidth]{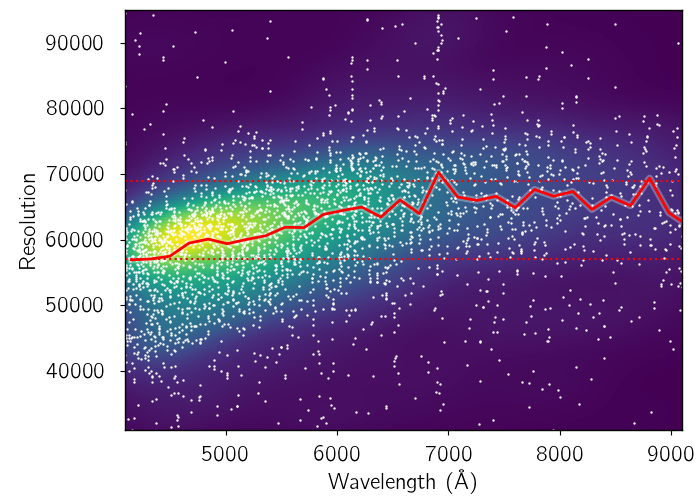}
\caption{{Resolution measured on $\sim3400$ ThAr lines across the whole wavelength range. The color code represents a density map of the individual measurements, represented as white dots. The average per order is shown by the red line.}}
\label{fig:res}
\end{figure}

The configuration of the new grating was selected to be the same than the original one. The only difference was a vertical flip of the grating that was originally up-side-down while it now faces up. {The overall response of the instrument remains very similar to the former configuration in terms of sensitivity and resolution as shown in Table~\ref{tab:CAFEprop}. The resolution was measured on $\sim3400$ ThAr lines across the whole wavelength range. The results show a mean resolution of $R=62\,000$ with lower values at blue wavelengths and higher at redder orders, resulting in a final resolution of R=$62\,000\pm5\,000$, as shown in Fig.~\ref{fig:res}.}

\begin{table}
\centering                                      
\setlength{\extrarowheight}{1pt}
\caption{Spectral properties of CAFE and \cafet.}              
\label{tab:CAFEprop}      
\begin{tabular}{l  l l}          
\hline                        
   & CAFE & CAFE$_2$ \\
\hline
Wavelength range   		&   3960-9500\AA			   &  4070-9250\AA\footnotemark[1]  \\
Resolution   			&   	62000$\pm$5000		   &  62000$\pm$5000 \\
Limiting magnitude\footnotemark[2]   &   	$V=15$ 	   &   $V=15$    \\
Limiting RV precision	&	20 m/s			 & 8 m/s \\

\hline                                             
\end{tabular}

\end{table}
\footnotetext[1]{Wavelength coverage provided by the \cafex\ pipeline. The actual coverage of the instrument extends down to 3900\AA\, including the Ca H\&K lines. Future versions of the pipeline will extend the extracted coverage to this regime.}
\footnotetext[2]{Maximum point source magnitude to achieve $S/N=20$ at 550\,nm in one hour of exposure time.}

\subsection{Active temperature control system}

In September 2018, a new active stabilization temperature control system was installed in the instrument room to create a controlled temperature environment at the spectrograph. The whole system is arranged over three rooms as shown in Fig.~\ref{fig:tempcontrol}.  At the CAFE optical bench room, two air handles (AH)  comprised of a fan, air to water chiller and a heating resistance create a laminar flow that surrounds the spectrograph. In the next room, an air to water chiller pumps cool water (9C) into the AH. A total of four precision temperature sensors (model PT100) are distributed at different locations in the optical bench and together with a temperature controller (model PTC10 from Stanford Research systems) monitor the thermal environmental conditions of the room (see an example for one night in Fig.~\ref{fig:cafe_temps}). 

Finally, a Programmable Logic control (PLC) handles two PID controllers, for the cooling and heating procedures respectively. The system flows a cool air injection into the room of about two degrees below the expected temperature, so the heating resistances perform a fine-tuning until the final operating goal temperature of 14$^{\circ}$C is reached. 

\begin{figure}
\centering
\includegraphics[width=0.50\textwidth]{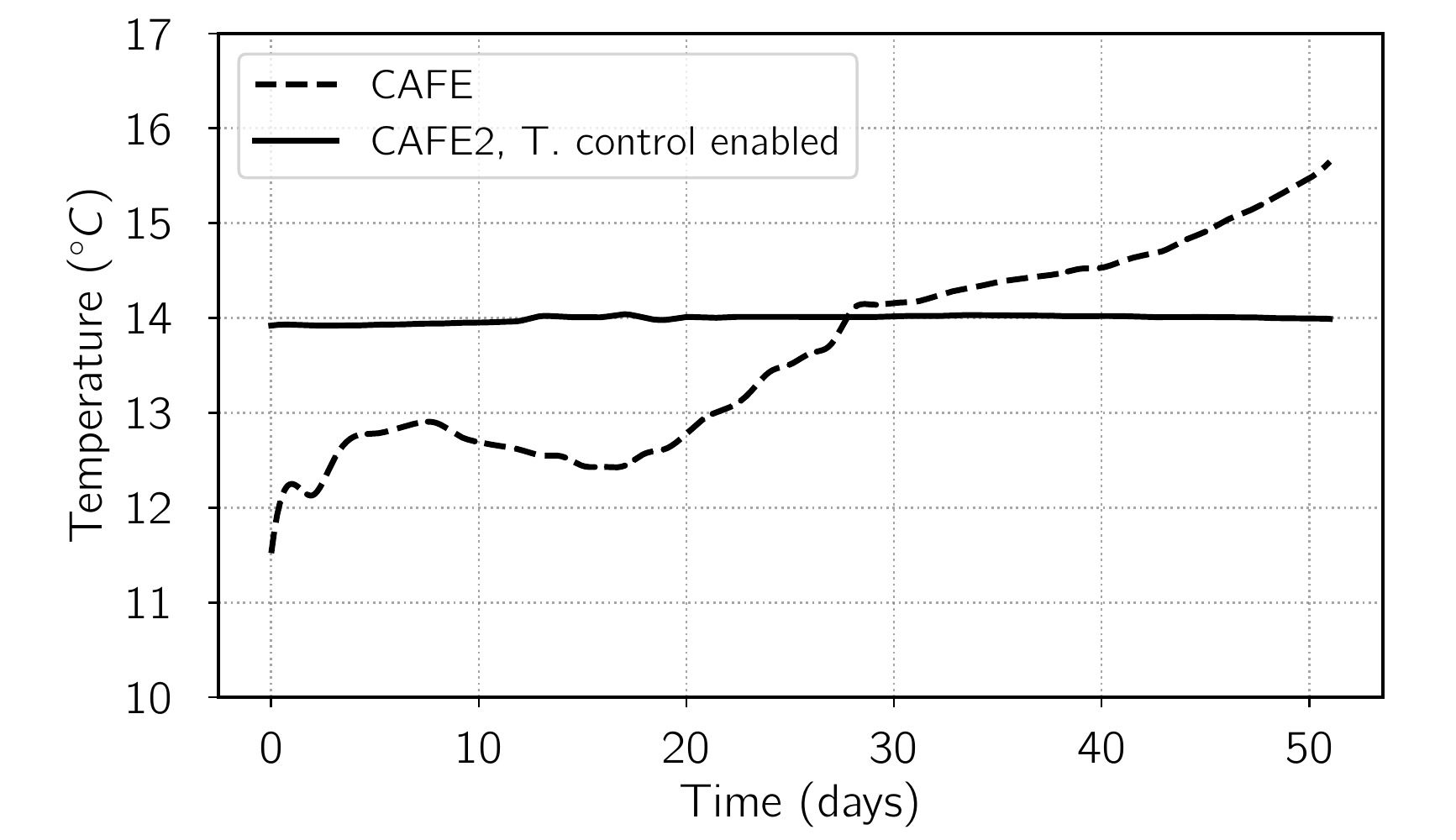}
\caption{Comparison of the temperature of CAFE room for 54 days with no temperature control (blue line) and the system enabled (red line). The standard deviation of the red line is about 0.02$^{\circ}C$. }
\label{fig:roomtemp}
\end{figure}

Figure~\ref{fig:roomtemp} shows the final performance of the system with respect to the original situation. Data were collected (with the room closed) in two periods of 54 days when the temperature controlled was disabled and enabled respectively and with a sampling of one data per day. During the uncontrolled time range, the temperature increased from 11.4$^{\circ}$C up to 16.8$^{\circ}$C due to the natural evolution of the room's temperature. When the temperature control was enabled, the measurement for the same time span was $14.01\pm0.18^{\circ}$C. The intra-night dispersion of the temperature is 0.011$^{\circ}$C, equivalent to the 3T chamber of the ESPRESSO instrument in one night. 

\subsection{CAFExtractor: the new public pipeline}

The upgrade of the instrument also comes with a new Python-based observatory pipeline (\cafex) that provides the user with the reduced data once the night ends. Around 5 hours are needed to fully reduce each night on the observatory computers. Details on this pipeline are given in section~\ref{sec:pipeline}. \cafex\ will be run by the observatory personnel after each night starting in period 2019B (July-2019). The outcomes of the pipeline (including the reduced spectra and the high level data products) will be made public with the raw data after the proprietary period at the Calar Alto Observatory Archive based on the Spanish Virtual Observatory\footnote{\url{http://caha.sdc.cab.inta-csic.es/calto/}}.

\subsection{Other improvements}

Besides all these new improvements, additional modifications have been done on the operational side. First, the header keywords have been modified to take the upgrade into account and to add key information previously missing in CAFE. In particular, the Julian date keyword (\texttt{JUL-DATE}) now provides up to seven decimal figures (previously only one was provided). Additionally, there are now four keywords providing the temperatures of the CAFE room (\texttt{TEMP4}), grating's mount (\texttt{TEMP3}), telescope pier (\texttt{TEMP2}), and collimator (\texttt{TEMP1}). Another relevant improvement is the installation of a new fiber shaker device. The new system now produces a mechanical scrambling in two axis (as compared to the previous one with just one axis motion). The two movements are obtained from one single engine, but the gearing is different for each axis so that the two-mode cycle is long. The robustness of this mechanical device was also improved.

We also developed a set of routines that monitor the health of different parts of the instrument to detect potential fails or degradation well in advance. In particular, these tools measure the bias and readout noise every day the instrument is used and compare them to the previous values. The values must be kept within some thresholds, otherwise warnings are raised by the health check scripts. They also measure the location and intensity of ThAr spots in the 2D frame and the position of the orders in the central column of the CCD on the flat frames. These routines help monitoring any possible displacement of the orders (critical for the reduction process and the stability of the instrument) as well as the health of the ThAr lamp, which ages every $\approx$ 230 hours turned on.

\section{\textit{CAFExtractor}: the new pipeline}
\label{sec:pipeline}

We have developed a new, fully automated, reduction and analysis pipeline for the instrument (called \cafex). Two versions of the pipeline have been delivered to the observatory, one for CAFE and one for CAFE$_2$. The pipeline version for CAFE will be used to provide high level data products of all instrument observations (including radial velocities for FGK stars) from its commissioning in 2012 to its shutdown in 2016. The results of this full reduction will be presented in a subsequent publication and the data products will be available through the CAHA archive based on the Spanish Virtual Observatory. In this section, we explain the basic steps performed by the pipeline. It is important to highlight that \cafex\ is partly based on routines from the CERES pipeline\footnote{Publicly available at \url{https://github.com/rabrahm/ceres}.} developed by \cite{brahm16}. As such, we will refer to that paper for further details on particular steps.

\subsection{Preliminary considerations}

Before describing the pipeline steps, it is important to remark some considerations about  \cafex:
\begin{itemize}

\item \cafex\  is prepared to be used by the observatory to provide science-ready data products to the \cafet\ observers. This means that it is not thought to be an on-the-fly data tool because it uses every night calibration data to accurately reduce the data. Consequently, for instance, morning calibrations are needed by the pipeline.

\item \cafex\ is a public pipeline and can be dowloaded by the users to perform their own reduction. A setup file allows the user to tweak different parameters that can change the results depending on the science case and target types. 

\item Part of \cafex\ is based on the CERES pipeline \citep{brahm16}. Some specific functions have been reused from this public pipeline and adapted or improved when necessary to work with \cafet.

\item \cafex\ assumes that the observations were carried out with a standard series of calibrations and science frames as recommended in Sect.~\ref{sec:recommendations}. This is: set of evening calibration frames (normally 20 bias, 20 flats, and 20 arcs), science frames in between arc frames (for proper best-precision wavelength calibration), set of morning calibration frames (normally 20 bias, 20 flats, and 20 arcs). 

\end{itemize}

\subsection{Basic pipeline steps}

\cafex\ is organized in ten reduction blocks (hereafter RBs). These RBs are called by a main routine (\texttt{cafex.py}) at the corresponding step of the reduction.  A setup module (\texttt{CAFEx\_SetupFile.py}, hereafter called setup file) contains the basic parameters that can be tweaked by the user to perform their own reduction of the data. These parameters and their default values are shown in Table~\ref{tab:defparams}. The pipeline also includes a routine with specific CAFE functions (\texttt{CAFEutilities.py}) and a routine with global functions (\texttt{GLOBALutils.py}), the latter one basically re-used from the CERES pipeline (see \citealt{brahm16} for details on the functions inside this module).

\vspace{0.2cm} \noindent \textbf{RB01 -- data preparation:} This RB is in charge of sorting and organizing the data files. The raw data is duplicated into a new working directory (the original raw data is then kept untouched) as specified in the \texttt{redfolder} parameter of the setup file. The files are renamed according to their type and adding the date of observation and the object name in the case of science frames. Then the different frames are sorted by type (bias, flat, arc, science) and stored in datacubes. 

\vspace{0.2cm} \noindent \textbf{RB02 -- instrument checks:} In this routine, the pipeline checks the quality of the data in order to find possible errors (e.g., flats with low number of counts, bias with too much counts or misclassified frames).  Additionally, it determines the shifts of a selected number of spots (around 100 spread around the whole frame) from each ARC file and compares them to a pre-selected reference file (parameter \texttt{RefArc} in the setup file). It cross-correlates the 2D frame of the first ARC of the night with the reference frame to estimate the global night shift in the X and Y directions. Then, it measures the shifts of the other ARCs of the night against the first ARC. The same is done for the intensity of the spots. A plot is stored in the \texttt{auxiliar} folder (see example in Fig.~\ref{fig:arc_shifts}). The  XY shift of the frame is computed through a 2D Gaussian fit to the cross-correlation of the two frames. If shifts are larger than 1 pixel, then the pipeline displays a warning message in red color. The temperature stability is also checked by displaying the temperature of the telescope dome, CAFE room, collimator, and grating for each frame as a function of time (see example in Fig.~\ref{fig:cafe_temps}). The pipeline prints in the terminal the delta and variance of these temperatures. Variations smaller than 20 mK are expected in operational status of the instrument.

\vspace{0.2cm} \noindent \textbf{RB03 -- master frames:} In this routine, the pipeline computes the Master frames that will be used to calibrate the science data. It computes Master frames for the different blocks of frames obtained during the night. Usually, this means two master biases, two master flats, and two master arcs (one for each twilight). The frames are clustered by Julian date based on a \textit{MeanShift} algorithm. The bandwidth for this algorithm can be modified in the setup file. The default is 1 hour for bias and flats, and 30 min for the arcs. A minimum number of frames to create a Master frame is necessary and this is defined in the setup file (default being 5 bias, 5 flats and 8 arcs). The frames inside each cluster are subsequently combined. BIAS frames are first clipped (with sigma-clipping value defined in the setup file, default being 5$\sigma$) and then averaged. FLAT frames are first clipped, then the closest MasterBias is removed and then the median of the clipped frames is taken as the MasterFlat. The ARC frames are again clipped, bias subtracted,  and averaged. Finally the SCIENCE images are bias subtracted with the closest MasterBias.

\vspace{0.2cm} \noindent \textbf{RB04 -- order tracing:} In this routine, the pipeline traces the orders to be extracted. This is done based on a reference location of the orders in the ReferenceFlat frame (defined as \texttt{RefFlat} in the setup file). The order tracing is done through a polynomial fit for each order. The order and amplitude of the polynomial are defined in the setup file as \texttt{order\_aperture\_ampl} (default$=5$) and \texttt{order\_trace\_degree} (default$=4$). For details on how the orders are identified please refer to section 2.2.3 in the CERES pipeline paper \citep{brahm16}.  The orders of the ReferenceFlat frame are searched for and traced. In a subsequent critical step, only some of the orders are selected according to previous analysis on the significance of their detection. The first order (starting from the bottom of the CCD) is then selected based on the nominal value from the ReferenceFlat. This value is set in the SETUP file as \texttt{y0Nominal\_first} and corresponds to the Y-pixel at $x=1024$ of the first order to be used (order 0). This number was computed manually on the ReferenceFlat and then refined with a first reduction of the night folder of the ReferenceFlat. The total number of orders selected is fixed by the setup file keyword \texttt{Nominal\_Nord} (default for \cafet\ is 79). Then the pipeline performs the order tracing on the MasterFlat frames using the y-shifts calculated in RB02, computes the root-mean-square (rms) of the fit (usually around 30 mpix), and checks and selects the nominal orders for \cafet\ in a similar fashion as for the ReferenceFlat.

\vspace{0.2cm} \noindent \textbf{RB05 -- order extraction and background:}  In this routine, the pipeline deals with the order extraction and the removal of the scattered light. First, it creates maps of scattered light for each MasterFlat (see section 2.2.4 in \citealt{brahm16}) and removes this scattered light from the MasterFlat. Then it uses this background-removed frames to extract the orders based on the tracing from the previous RB04. Finally, the MasterFlat is normalized using the module \texttt{FlatNormalize\_single} from the CERES pipeline. A similar procedure is subsequently applied to the ARC, MasterArc and SCIENCE frames, this time also dividing the extracted frames by the normalized flat. 

\begin{figure}
\centering
\includegraphics[width=0.5\textwidth]{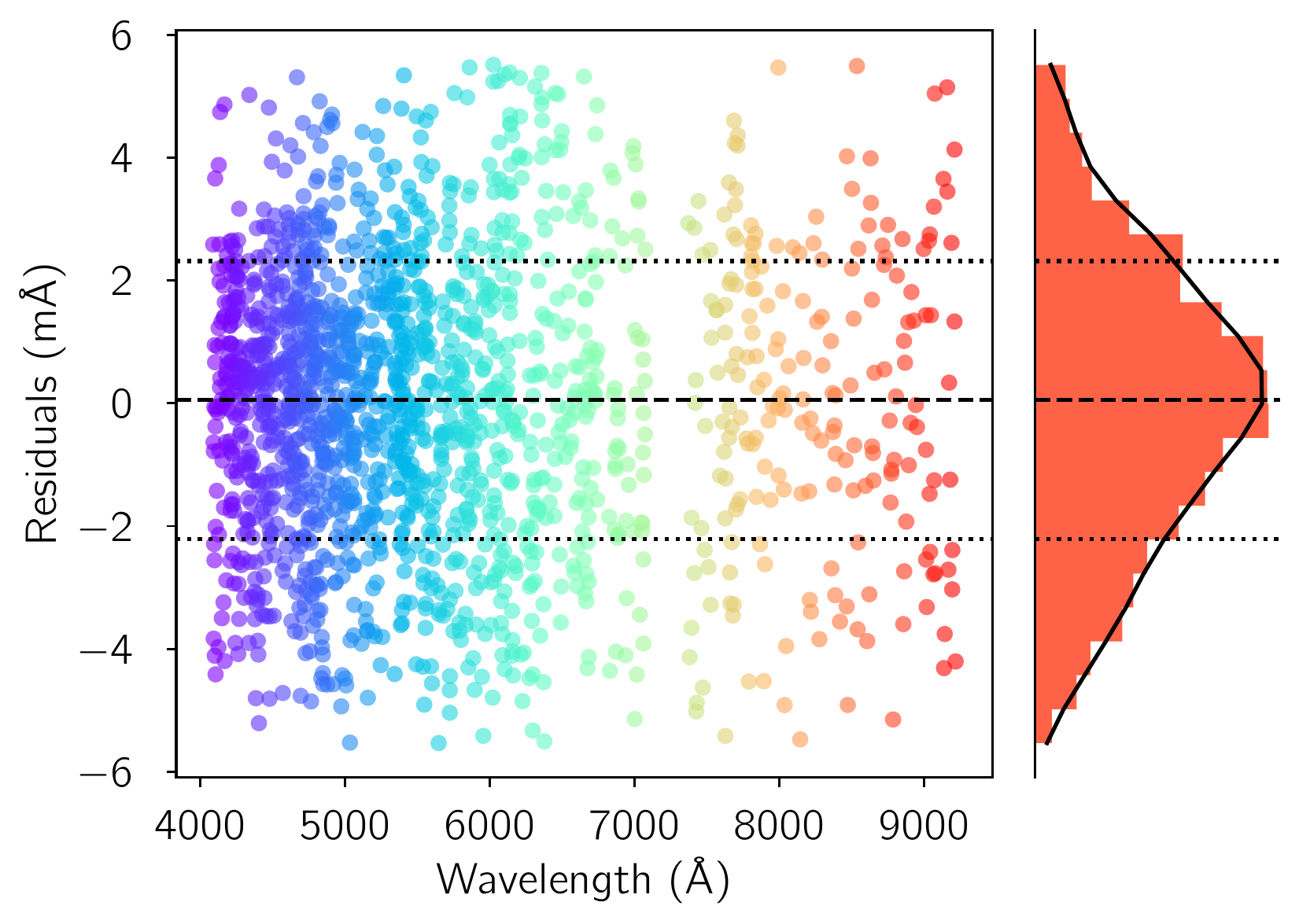}
\caption{Residuals of the wavelength calibration of a MasterArc for all ThAr lines used in the process. A typical dispersion of 2 m\AA\ is found. The dotted lines correspond to $\pm1\sigma$. The right panel shows the density distribution of these residuals.}
\label{fig:wavecal}
\end{figure}

\vspace{0.2cm} \noindent \textbf{RB06 -- wavelength calibration:} This is a critical block of the pipeline. The wavelength calibration (WC) is performed by following the equations from the CERES generic pipeline (see section 2.4.1 in \citealt{brahm16}) but the procedure has been modified to optimize the number of lines to be used in the calibration process. For each MasterArc, the pipeline calculates the pixel shift ($\Delta$) of the current frame with respect to the nominal pixel values. The nominal pixel values are stored in ascii tables named \texttt{cafex\_order\_X.dat}, where X corresponds to the physical order. This is done through the cross-correlation of each extracted order with a binary mask including the ThAr lines in these tables. These lines were previously selected to be isolated and strong enough to be used in the wavelength calibration process. This $\Delta$ is then used to measure the pixel centers of the ThAr lines specified in the stored files. A region of $\pm$5 pixels around each ThAr line is selected and fitted with a 1-Gaussian profile (1G) and a 2-Gaussians (2G) profile to account for possible blended lines. The Bayesian Inference Criterion (BIC) of each model is compared and only if BIC$_{2G}$ < BIC$_{1G}$ then the 2G model is selected. Additionally, only the well-fitted lines ($\chi_{\rm red} < 5$) are used to compute the wavelength solution. The pixel center, corresponding wavelength, and order number for each ThAr line are then stored and passed to the wavelength solution computation. In a first step, an order 3 polynomial is fitted and a 3$\sigma$ clipping is applied. The process is repeated over 5 iterations. Note that only if the number of valid lines at this step is larger than 7, the order will be used in the global wavelength fitting. The global wavelength solution is computed based on the lines that survived the previous step. A 2D Chebyshev polynomial is fitted based on Equation 3 in \cite{brahm16} by using the \texttt{Fit\_Global\_Wav\_Solution} function of the CERES pipeline. The order of the polynomial is determined in the SETUP file (default is $n_x=3$ for the pixel direction and $n_m=8$ for the order direction). {A typical 2-3 m\AA\ precision is achieved on the ThAr lines finally used. This corresponds to ~100-120 m/s per spectral line or  2.5-4 m/s assuming the total number of ThAr lines finally used, typically around 1700 lines), see Fig.~\ref{fig:wavecal}. We note, however, that this is an optimistic estimation of the error budget from the wavelength calibration as it corresponds to the exact pixels (ThAr lines) used to compute the wavelength solution. A more global precision of the wavelength calibration was measured on an individual night by obtaining the wavelength calibration of 5 individual ThAr frames and measuring the wavelength dispersion of each pixel along these frames. The median dispersion of the pixels in the CCD (after removing the first and last orders and the 300 first and last pixels to avoid edge-effects) is 5.5 m/s, which we consider as the error budget from the wavelength calibration process of CAFE}. The final output of this RB is a wavelength calibration matrix for each of the MasterArcs of the night. 


\begin{figure}
\centering
\includegraphics[width=0.5\textwidth]{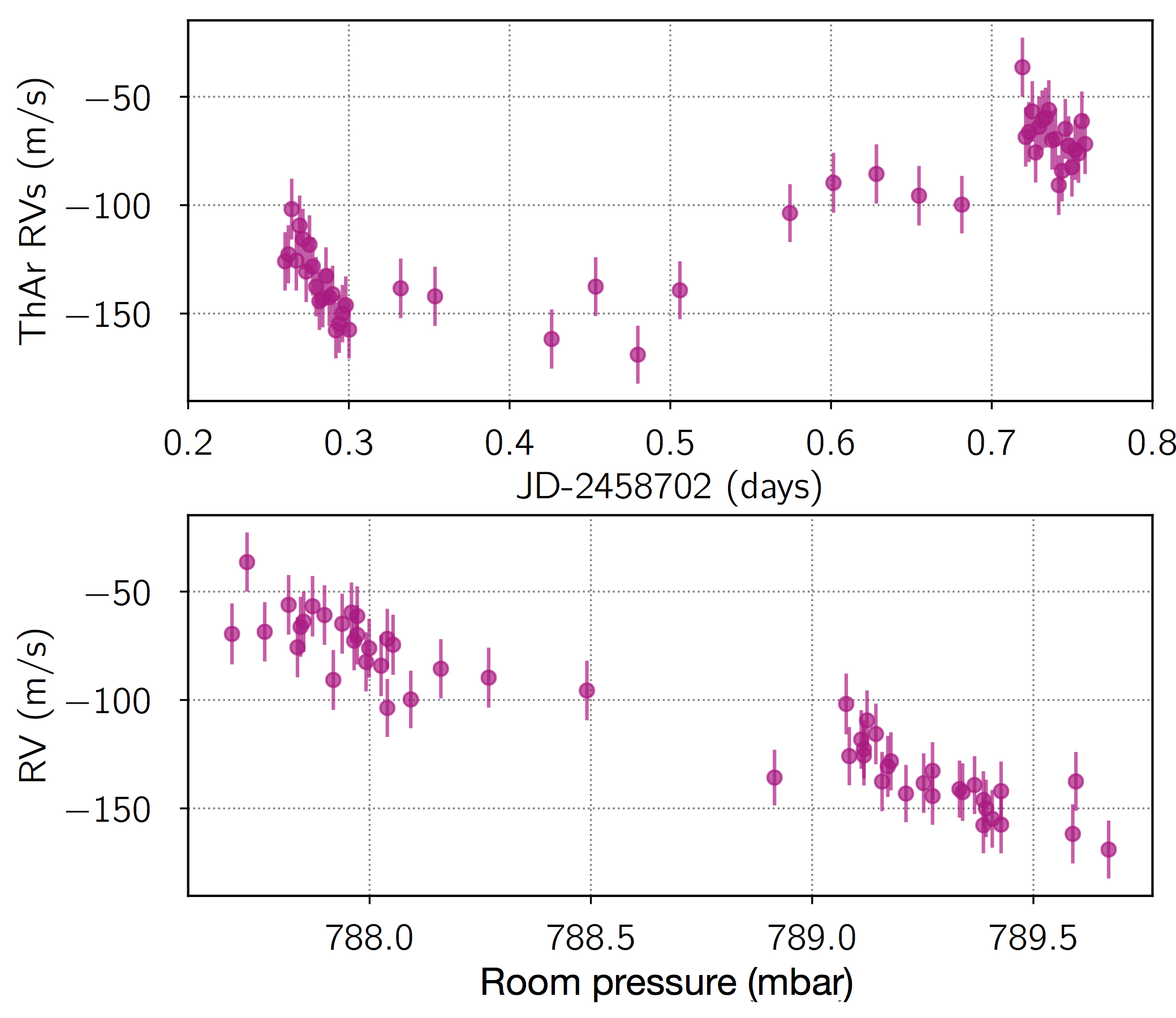}
\caption{\textbf{Upper panel}: Radial velocity measurements on the arc frames during a normal observing night. Top panel shows the whole dataset from the evening calibrations (symbols clustered to the left), the reference frames obtained along the night, and the morning calibrations (symbols clustered to the right). The night drift is evident. \textbf{Bottom panel}: Dependency of the measured radial velocity on the ThAr frames with the room pressure.}
\label{fig:arcRVs}
\end{figure}

\vspace{0.2cm} \noindent \textbf{RB07 -- application of WC to arc and science data:} Once we have a master wavelength calibration for the night, {we measure its radial velocity (v$_{M}$) through a cross-correlation with the ThAr mask. Ideally, this RV should be around zero and represents the zero point radial velocity of the night. We then proceed to apply the master wavelength solution to the science frames. In a first step, the wavelength solution of the Master frame is applied to all individual ThAr frames obtained along the night. Then, we measure the radial velocity of each of the ThAr frames (v$_{\rm ThAr,i}$) again through a cross-correlation with the ThAr mask. These values usually span a range of velocities of around 100-150 m/s with a clear trend along the night (see upper panel in Fig.~\ref{fig:arcRVs}) due to ambient pressure variations (see Sect.~\ref{sec:stability}). In the absence of a second fiber or an iodine cell, these variations can be corrected with ThAr frames obtained close in time to the science frames. Once the velocity of the individual ThAr frames are measured, we calculate the velocity correction for each science frame $j$ ($v_{\rm corr,j}$) through a linear interpolation in time between the velocity measured for the closest arcs before and after the science frame (ideally, this would have been taken right before and after the science frame). This correction is then applied to the wavelength solution to get the final wavelength solution for the science frame:
 }
 
\begin{eqnarray}
\lambda_j = \left(1-\frac{v_{\rm corr,j}-v_{M}}{c} \right) \lambda_{M}
\end{eqnarray}

\vspace{0.2cm} \noindent \textbf{RB08 -- radial velocity:} In this routine the pipeline calculates radial velocities of the science frames whenever possible, using cross-correlation technique against a binary mask \citep{baranne96}. This binary mask is taken from the HARPS instrument \citep{mayor03}. For the wavelength range not covered by HARPS, we use a rebuilt set of lines based on the CERES binary mask. The cross-correlation function (CCF) is only built from a particular number of orders explicitly selected to avoid those regimes with a high number of telluric lines or containing few information from the binary mask. Additionally, the first and last 300  pixels from each order are removed from the calculations to avoid edge effects and the order overlap, especially in the blue region. A first guess of the RV is estimated by constructing a temporary CCF with 4 orders spread around the spectrum with a broad radial velocity coverage from -200 to 200 km/s. The velocity array is constructed in steps of 0.25pix in velocity space (corresponding to roughly 500 m/s). This temporary CCF is then fitted with a Gaussian profile to estimate the RV. A new velocity array is then constructed around the guessed RV with a $\pm$ 20 km/s amplitude in case the estimated width of the CCF ($\sigma_{\rm CCF}$) is below 10km/s. Otherwise the amplitude of the velocity array is set to 5$\times \sigma_{\rm CCF}$. All selected orders are then used to construct the CCF. The final RV is estimated as the center of a Gaussian profile fitted by the pipeline to the final CCF (result of summing up the CCFs of all selected order). 
The information from the individual CCFs and the final profile is stored and saved in the header of the finally reduced fits file. The pipeline also generates plots of the CCF analysis as shown in the example Fig.~\ref{fig:ccf}.

\vspace{0.2cm} \noindent \textbf{RB09 -- normalization and merging:} 
The extracted SCIENCE frames are normalized order by order using a simple order 2 polynomial. The S/N per pixel of each order is also estimated in the same step by using the \texttt{derSNR} algorithm \citep{stoehr08}. The spectrum S/N is provided as the S/N corresponding to the order containing the 550nm wavelength (order 41 in current status of \cafet). The normalized orders are then merged into a single array. The overlap between the orders is treated as follows. The first and last 200 pixels from each order are trimmed. In overlapping order cases, the mean of the two orders after interpolating to a common wavelength array is obtained and saved. Hence, we warn that the merged spectrum is not suitable for high-precision radial velocity measurements. In that case, we encourage the use of the individual orders. 

\vspace{0.2cm} \noindent \textbf{RB10 -- results saving:} 
The final step corresponds to the saving of the reduced files. This process includes the creation of the new header including the raw image header information and adding additional products obtained by the pipeline (see next section). The reduced files are then saved in a \texttt{reduced} folder inside the one specified by the user in the setup file.

\subsection{Pipeline data products}

\cafex\ provides the user with wavelength calibrated, normalized and 1D-merged spectra. Additionally, it performs precise estimations of the RV for FGK stars through the CCF analysis as explained above. The main products of the pipeline are the reduced files, included in the \texttt{reduced} folder. These FITS files contain nine extensions as described in Table~\ref{tab:fitsformat}. The header contains the keywords from the raw image plus additional information obtained during the pipeline process. 

Additionally, a series of diagnostic plots are stored in the \texttt{auxiliary} folder. In particular, a plot showing the arc frames XY shifts and relative intensity variations during the night, a plot displaying the temperatures of the instrument sensors along the night, one plot per ARC and SCIENCE frame showing the CCF and individual results for each order (RV, FWHM and CCF height), and a plot showing the RV drift measured from the ARC and MasterArc frames along the night.

\section{Instrument and pipeline performance}
\label{sec:performance}

\subsection{Stability of the reference frames}
\label{sec:stability}

Despite of the stabilization of the room temperature, {CAFE is still not stabilised in pressure (as opposed to the temperature). A new sensor has been installed inside the CAFE room to monitor these pressure variations along the night. Typical intra-night variations in pressure are of the order of 1-2 mbar. These changes modify the profile of the spectral lines, which translates into velocity trends along the night (see bottom panel in Fig.~\ref{fig:arcRVs}). This variation (increasing up to several hundreds of m/s depending on the temperature gradients of the night) still makes it necessary to take reference frames before and after each science frame in order to properly correct for this drift during the wavelength calibration step.}

\subsection{Radial velocity stability and precision}
During the commissioning runs, we observed different radial velocity standards with the purpose of testing different effects. 

\subsubsection{Dependency with signal-to-noise ratio}

\begin{figure}
\centering
\includegraphics[width=0.5\textwidth]{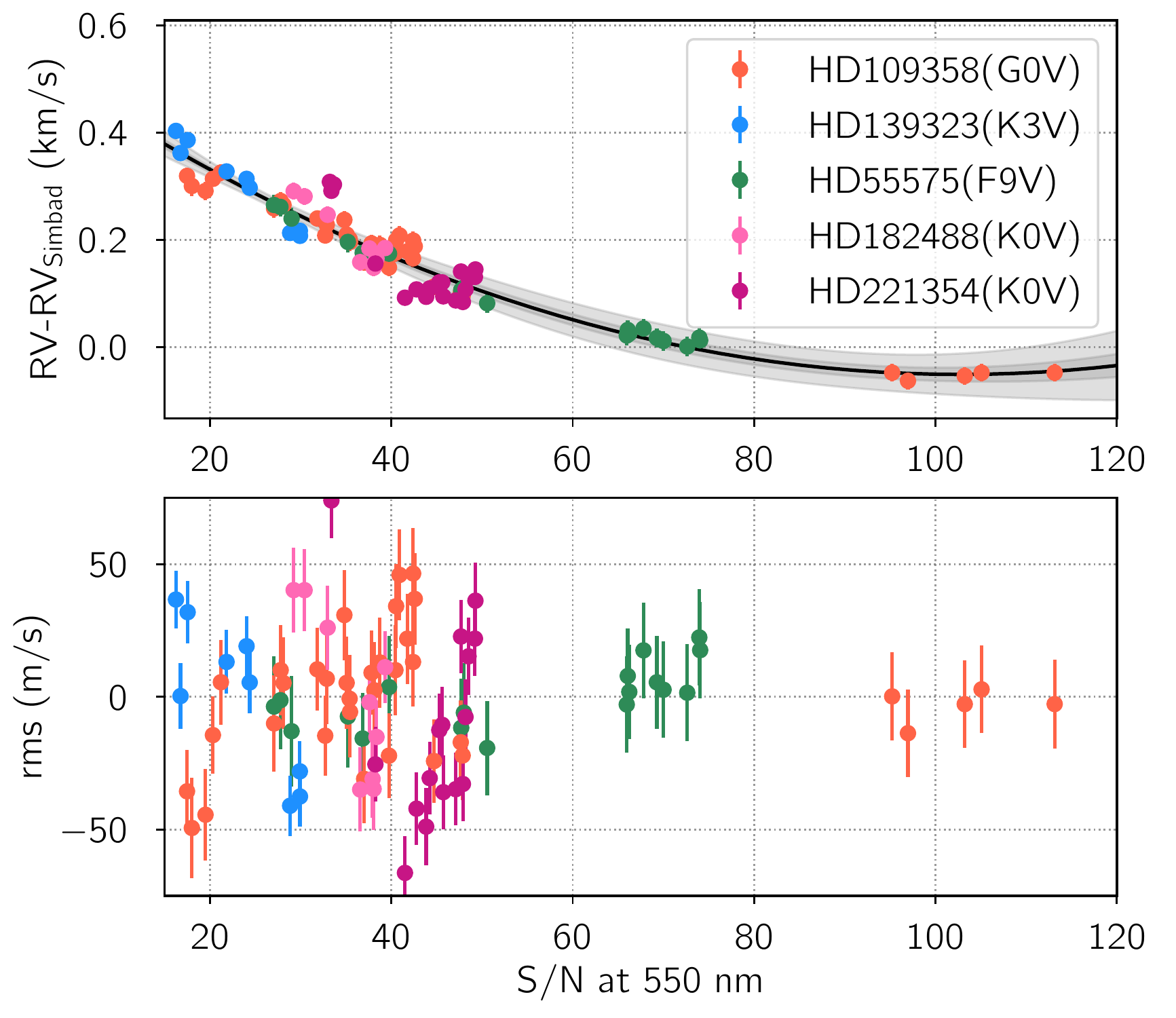}
\caption{Signal-to-noise effect on the radial velocity in the CAFE instrument for different radial velocity standards. The upper panel shows the measured radial velocity compared to the true radial velocity as published in Simbad versus the S/N at 550 nm. The black line represents the best second-order polynomial fit and the shadowed regions correspond to $1\sigma$ and $3\sigma$ confidence levels. The lower panel shows the corrected values of the radial velocity, displaying a flat dependency with the S/N.}
\label{fig:snreffect}
\end{figure}

We observed the radial velocity standard stars   HD\,139323 (K3V, $V=7.6$ mag), HD\,55575 (F9V, $V=5.6$ mag), and HD\,109358 (G0V, $V=4.3$ mag) with different exposure times and under different seeing conditions. This corresponded to a S/N range for the observations of these three standards between S/N$=1$ to S/N$=120$. Their measured radial velocity displays a clear dependency with the S/N of the spectrum in this range (see upper panel of Fig.~\ref{fig:snreffect}). This dependency can be as large as 500 m/s in spectra of the same target obtained with a wide range of S/N. {The origin of this dependency is currently unknown. We can discard the charge transfer inefficiency effect because the CCD is read out in the spatial axial, so perpendicular to the spectral direction. Consequently, its effect has no direct implication on the contiguous pixels in the wavelength direction. We also did not find a correlation of the radial velocity with the seeing effect as described in \cite{boisse10a}.}

Fortunately, the effect seems not to depend on the spectral type of the target (at least in the FGK range) and so we can determine a correction factor for the radial velocity given the commissioning data. We found that a second-order polynomial appropriately corrects for this effect. Once corrected, a flat root-mean-squared (rms) of 8 m/s (scatter of the RV measurements) can be achieved for spectra with S/N$>50$ and a 19 m/s rms if we account for spectra with S/N$>20$  (see lower panel of Fig.~\ref{fig:snreffect}). 

\subsubsection{Fiber centering effect}

In this test, we aim at measuring the quality of the double-direction mechanical scrambling applied to the fiber of the instrument. In this case, we obtained a series of spectra of the same radial velocity standard (HIP\,70873) by positioning the target in different locations of the fiber hole. In particular, we performed the sequence: 3xC, 3xN1, 3xN2, 3xC, 3xS1, 3xS2, 3xC, 3xE1; where C, N, S, E refer to Center, North, South and East respect to the center of the fiber, respectively. Also, "1" or "2" refer to the amplitude of the offset , with "1" corresponding to 0.6 arcsec from the center and "2" corresponds to 1.2 arcsec. We obtained three spectra of 300\,s at each position to check for possible variations unrelated with this test. 

\begin{figure}
\centering
\includegraphics[width=0.5\textwidth]{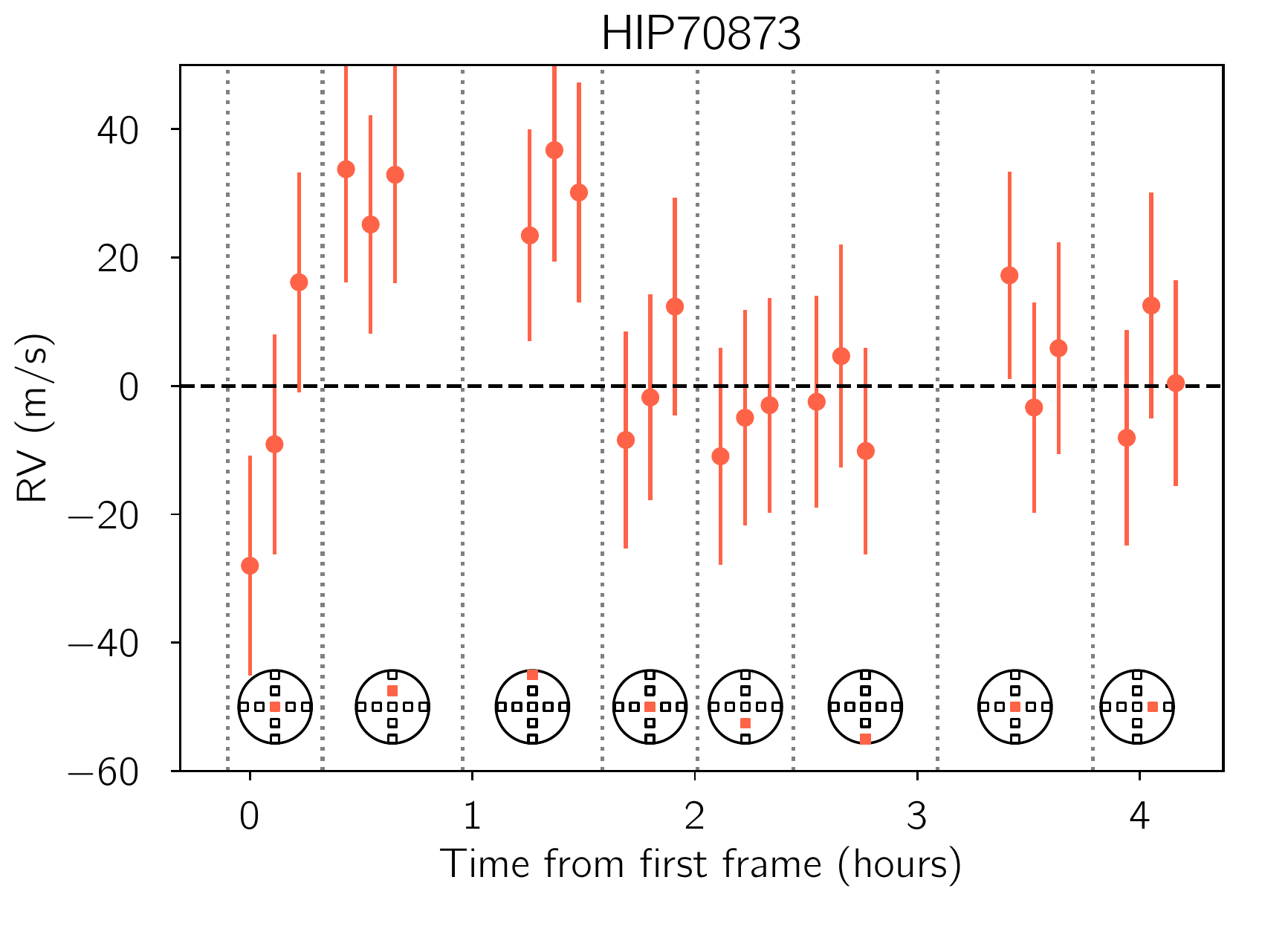}
\caption{Test for the decentering effect on the radial velocity of CAFE. The radial velocity series was performed by locating the star at different locations inside the fiber hole. These locations for each set of 3 measurements is indicated in the bottom part of the figure, where the circle represents the fiber hole and the filled square represents the location of the star for those measurements. }
\label{fig:fiber_centering}
\end{figure}

The results are presented in Fig.~\ref{fig:fiber_centering}. The radial velocities are corrected from the SNR-effect previously described and the exact location where the star was positioned is indicated in the bottom part of the plot. As shown, only positioning the star Northwards of the fiber center seems to have an effect in the radial velocity measurement, with a 30 m/s jump with respect to the other positions.  This effect needs further investigation. However, it is important to mention that it would only be relevant in cases of very bad (and unlucky) guiding conditions with the star drifting northwards. In any other case, the scrambling of the fiber is good enough to keep the radial velocity precision at the 8 m/s level, as it is the case of the last 15 measurements of this test.

\subsubsection{Seeing/Focus effect}

We performed a series of observations of the radial velocity standard HD\,55575 changing the focus of the telescope to simulate and check the influence of varying seeing conditions. The results of this test show that there is no additional effect besides the loss of flux and hence of S/N, which introduces the SNR-effect previously described. After correcting for this effect, no additional trend is detected in this dataset above 8 m/s.

\subsection{Throughput}
\label{sec:throughput}

\begin{figure}
\centering
\includegraphics[width=0.5\textwidth]{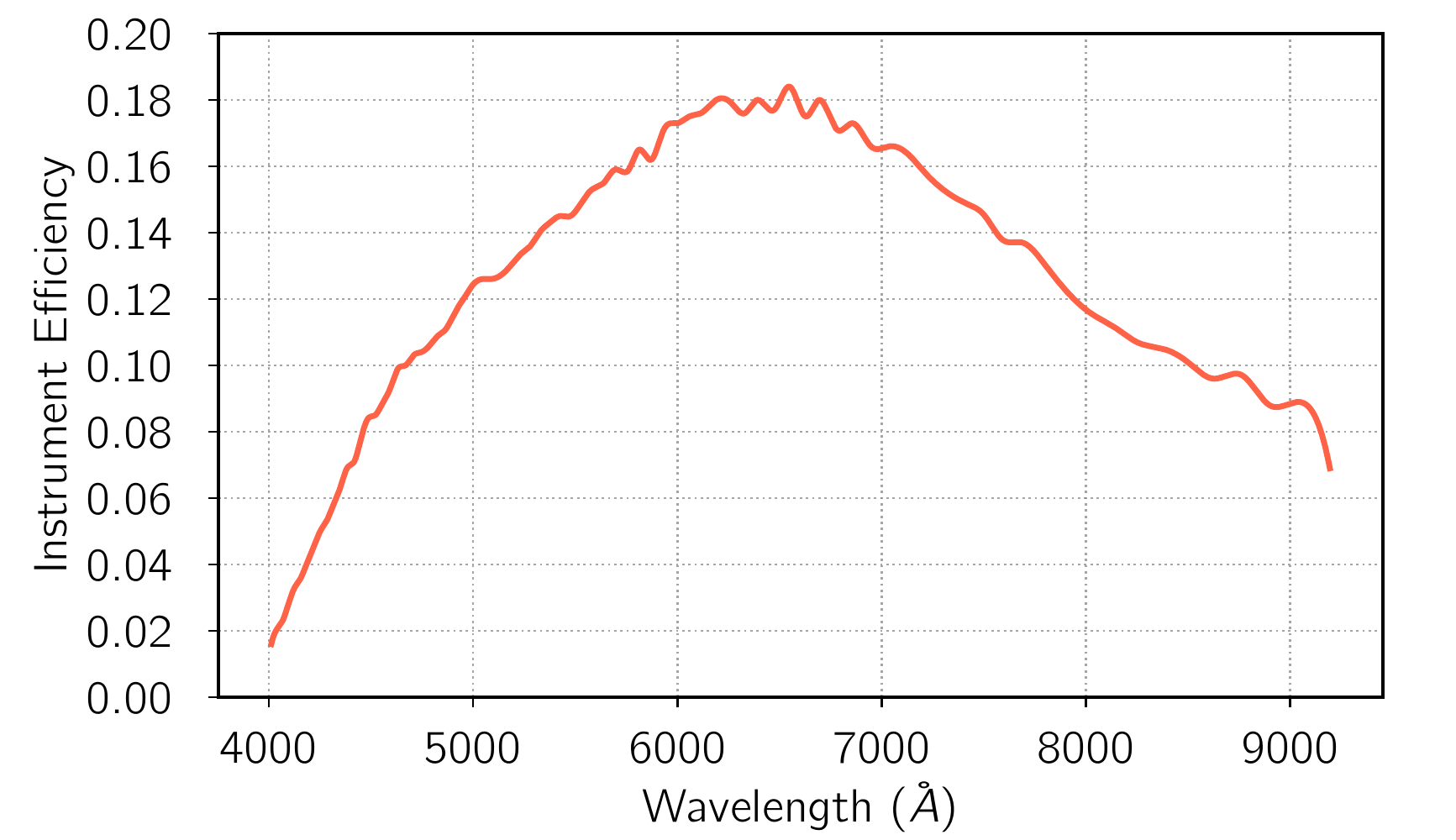}
\caption{Throughput of the \cafet\ system including fiber, spectrograph, and CCD (see Sect.~\ref{sec:throughput}).}
\label{fig:throughput}
\end{figure}

{The efficiency of CAFE, after the grating replacement, was derived from observations of the flux standard BD+25d4655 on 25-July-2018. We observed this star under good weather conditions (1.5 arcsec seeing and 0.19 mag extinction). We estimate the throughput of the telescope, fiber, spectrograph, and CCD system as the ratio between the detected number of photoelectrons (N$_{\rm e^-,det}$), and the expected value from the star (N$_{\rm e^-,exp}$) as measured by \cite{oke90}. We measured N$_{\rm e^-,det}$ from the counts per pixel in each order of the extracted spectra and transforming this into photoelectrons using the 1.2$e^-$/count gain selected in the Andor-iKON-L DZ936NBV camera. The N$_{\rm e^-,exp}$ was calculated from the flux density provided in \cite{oke90} and \cite{hamuy94}. In those catalogues we find the flux of this star being $10^{-16}$ erg s$^{-1}$ \AA$^{-1}$ cm$^{-2}$. In the case of \cafet, the flux is collected by the effective area of the 2.2m telescope (3.183 m$^2$) in a wavelength interval $\Delta\lambda$ given by the dispersion of each order and divided by the energy of a single photon of that wavelength. N$_{\rm e^-,exp}$ is reduced by the atmospheric extinction, measured in the V band by the extinction monitor in the observatory (\texttt{CAVEX}) and calculated for each wavelength using the theoretical extinction curve \citep{hayes75} for Calar Alto Observatory. Also reduced by the light losses at the entrance of the 200$\mu$m (2.34 arcsec) diaphragm that depends on the seeing measured by the Calar Alto seeing monitor (RoboDIMM) and at the 100$\mu$m width slit. The reflectivity of the primary (0.82) and secondary mirror (0.85) measured with a Iris908RS reflectometer is also taken into account. Fig.~\ref{fig:throughput} shows the derived efficiency as a function of wavelength for the system including fiber, spectrograph, and CCD.}

\subsection{Science performance}

\begin{figure*}
\centering
\includegraphics[width=1\textwidth]{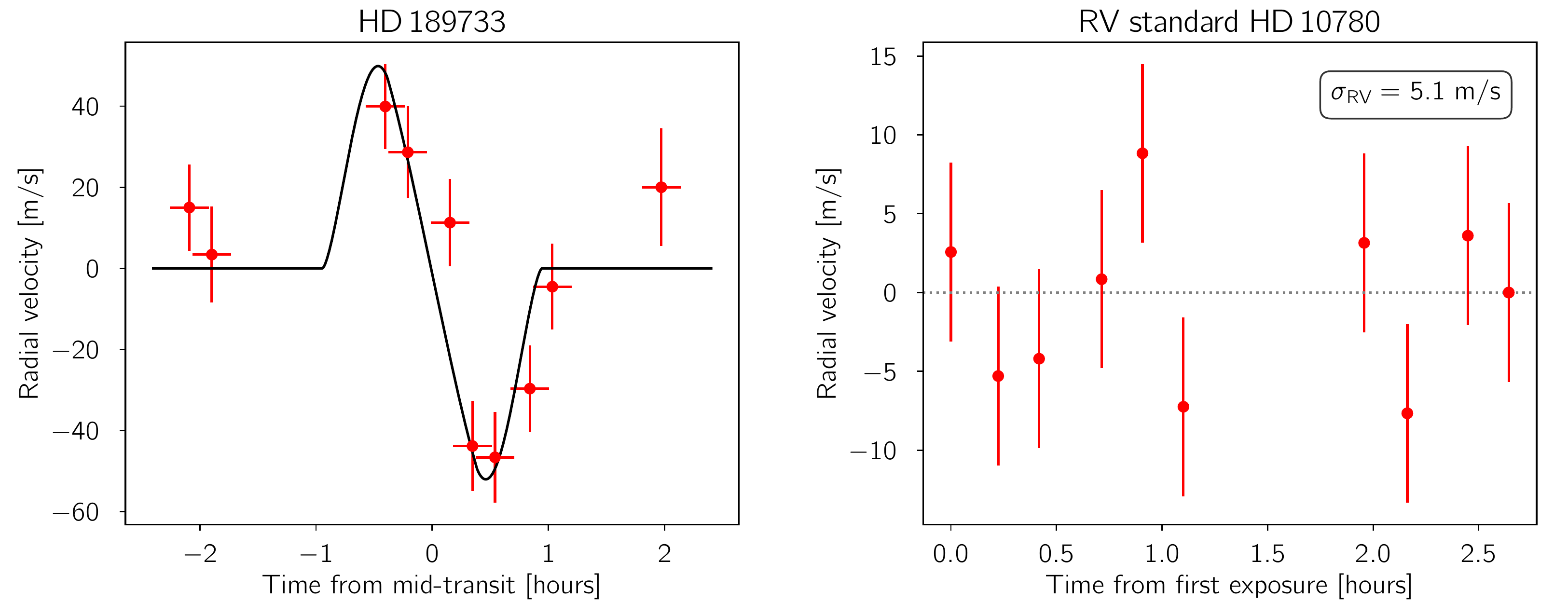} 
\caption{{Science performance of \cafet. \textbf{Left:} Rossiter-McLaughlin effect of the extrasolar planet HD\,189733\,b seen by \cafet on the night of September 19th 2019. The black line represents the model based on the precisely known parameters from (\citealt{digloria15}). \textbf{Right:} Radial velocity time series of the standard star HD\,10780 (\citealt{udry99}) along 2.6 hours of continuous observations. The rms of the 10 exposures is 5.1 m/s. The S/N of the spectra is around 60 at 550\,nm.}}
\label{fig:sci_results}
\end{figure*}

{As a final step in the validation process of the upgraded \cafet, we performed  two tests to verify the scientific value of the instrument. First, on the night of September 18th 2019 we observed the transit of the well-known hot-Jupiter HD\,189733\,b \citep{bouchy05} with the aim of detecting the Rossiter-McLaughlin effect induced on the radial velocity of the star due to the planet occulting part of the stellar disc along the transit. This effect was previously measured in this planet with high precision by \cite{digloria15} using the HARPS instrument. The transit duration is 1.8 hours and the stellar magnitude is 7.7 mag in the V band. We obtained 10 spectra of 10-min exposure time of this target, 2 of them before the transit, 7 in transit, and one after the transit. The data were reduced by \cafex\ as in a normal night. The results of these observations are shown in Fig.~\ref{fig:sci_results} (left panel), where we have overplotted the Rossiter-McLaughlin model according to the parameters published in \cite{digloria15}. The result shows a clear agreement with the expected effect, with a 40 m/s semi-amplitude.}

{The second test consisted on obtaining a radial velocity time series of a standard star to test the stability of the instrument along several hours. We observed the radial velocity standard star HD\,10780 (V = 5.6 mag, \citealt{udry99}) with an exposure time of 10 min per spectrum (reaching a S/N of 60 at 550\,nm) and taking arc frames every 30 min (approximately). The data was reduced using the \cafex\ pipeline and the resulting RVs are shown in Fig.~\ref{fig:sci_results} (right panel). The rms of the radial velocities in this dataset is 5.1 m/s, which demonstrates the best capabilities of CAFE under stable conditions and along several hours.}

\begin{figure}
\centering
\includegraphics[width=0.48\textwidth]{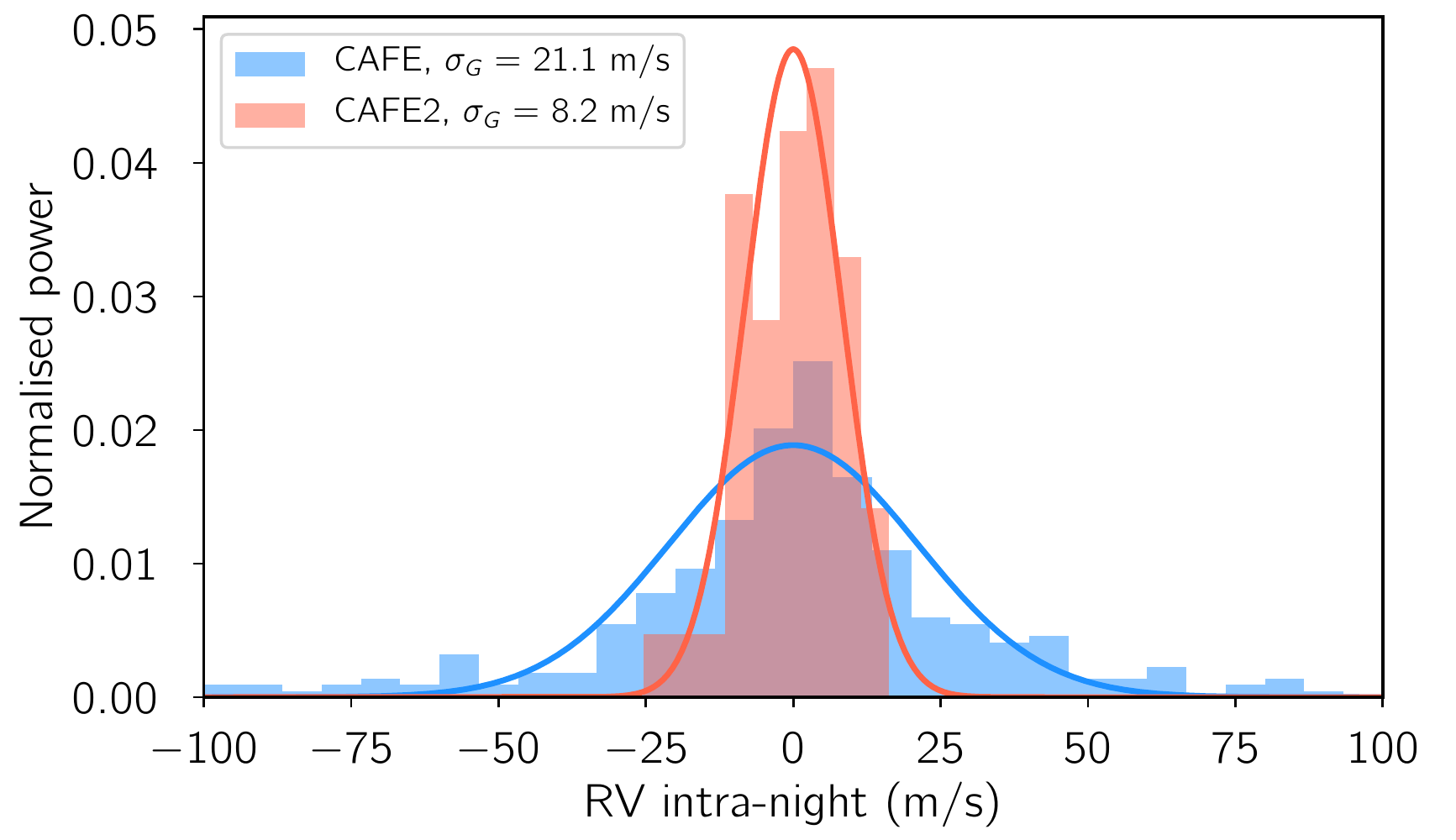} 
\caption{Radial velocity precision before and after the CAFE upgrade. Comparison of the intra-night radial velocities for the same standard star including all nights when the standard was observed at least 5 times along the night. The dispersion for CAFE and \cafet is 21.1 m/s and 8.2 m/s, respectively. }
\label{fig:CAFEcomp}
\end{figure}

\subsection{Observing recommendations}
\label{sec:recommendations}

Given the performance of the instrument and pipeline, in case of programs requiring radial velocity precisions better than 500 m/s, the following strategy should be followed when planning and performing the observations:

\begin{itemize}
\item Observe a radial velocity standard per night to measure nightly zero points and to test the radial velocity correction due to the SNR-effect. A list of commonly used (well tested) RV standards with this instrument is published on the CAFE website\footnote{\url{http://www.caha.es/CAHA/Instruments/CAFE/}}. 

\item Obtain one ThAr reference frame before and after the science observation (or at least one every half an hour) to properly account for the night drift. As shown, despite of the active temperature control of the CAFE room, there is still a temperature trend in the grating of the system that could be correlated with the intra-night radial velocity drift measured from the ThAr frames. As a consequence, only having ThAr frames before and after the observations would allow reaching the limiting radial velocity precision of the instrument (i.e., below 10 m/s for $S/N>50$). It is important to remark that the drift does not depend on the position of the telescope so it is not strictly necessary to obtain a ThAr frame before and after each science frame but to ensure that a science frame is sandwiched between two arc frames. 

\item Long exposures ($>30$ min) will not provide a proper radial velocity correction for the night drift. Hence programs requiring high-precision radial velocity ($\sim 10$~m/s) must be restricted to FGK stars brighter than $V<13$ mag. 

\item The pipeline needs a set of evening and morning calibrations, requiring $>10$ frames of biases, flats and arcs. Although these are usually taken by the observatory personnel, it is always recommended to ensure that a proper number of calibration frames was taken close in time to the observations (at least few hours before). 

\end{itemize}

\section{Summary and conclusions}

We have presented in this paper the improvements performed to the CAFE instrument, including the installation of a system to actively control the temperature of the CAFE room and the replacement of the damaged grating by a brand-new one. The temperature inside the CAFE chamber is fixed at an operating temperature of 14$^{\circ}$C and is now stable during one night at the $\sim$ 20 mK precision. The new grating now provides a throughput similar to the original measurements of the instrument when it was firstly installed and commissioned in 2012. Added to this major improvements, other minor changes have also been implemented, specifically in the header of the fits files (now including a more precise Julian date and information about the temperature of the different sensors inside the instrument), a set of routines to control the health of the instrument to be run by the observatory personnel every night the instrument is used, and a new mechanical two-mode scrambler for the optical fiber.

A new Python-based pipeline, called \cafex\ (or simply CAFEx) has also been developed and installed at the observatory to fully reduce the CAFE data after each night. The pipeline performs the basic reduction steps (bias removal, flat-fielding, order tracing and extraction and wavelength calibration, oriented to high-precision radial velocity studies) and provides the extracted spectra in different formats: order by order, normalized and merged. The pipeline also provides high-level products like the S/N per order, the S/N at 550 nm, the radial velocity of the spectrum measured by cross-correlating the spectrum with a binary mask with more than 2000 lines obtained from a G2V star. It also accounts and corrects for the radial velocity drift by using the reference frames obtained during the night, and accounts for the S/N-effect of the instrument. 

With all these improvements, the performance of the instrument coupled with the new pipeline now provides a wavelength coverage of 4070-9250\AA\ (although a dedicated reduction and future versions of \cafex\ could extract bluer orders down to 3900\AA) and a radial velocity precision of 8 m/s for $S/N>50$ after correcting for the effects described in Section \ref{sec:performance}. This precision is an improvement of more than 100\% compared to the previous state of the instrument and previous pipeline, providing 20 m/s in the same regime. This precision now allows the detection and characterization of extrasolar planets down to the Neptune-mass regime for periods below 10 days. 

Being public, the new pipeline will provide the users with science-ready data immediately after obtaining the observations and will populate the CAHA archive\footnote{\url{http://caha.sdc.cab.inta-csic.es/calto/}} (managed by the Spanish Virtual Observatory) with raw and reduced data to be released after the proprietary period. 

As in any other instrument, there is still room for improvement to increase the stability and performance of \cafet. In particular, we identify different points regarding software and hardware that need to be addressed. From the hardware point of view, replacing the current circular fiber by an octagonal one will  improve the homogeneity in the far field and thus will reduce the dependency of the radial velocity with guiding or bad centering effects. Additionally, the instrument would also benefit from the installation of a simultaneous Fabry-Perot to more precisely measure the night radial velocity drift. From the software point of view, the guiding system needs to be tested for long exposures (longer than 30 min.), we need to extend the correction of the S/N-effect for high S/N spectra ($S/N>80$), measure nightly zero points (if there are any larger than the instrument precision) as in the case of CARMENES \citep{trifonov18}. In this sense, it is important to remark that the pipeline will be updated regularly to account for the potential instrument changes implying radial velocity jumps. The latest version and news can be downloaded from the Github repository\footnote{\url{https://github.com/jlillo/cafextractor}}, also accessible from the CAFE website\footnote{\url{http://www.caha.es/CAHA/Instruments/CAFE/}}.


\section*{Acknowledgements}
This work has only been possible thanks to the hard dedication of the Calar Alto Observatory personnel, involving all departments. Special thanks to all support astronomers for carrying out the observations, to the engineers (specially Daniel Ben\'itez, Julio Mar\'in y Juan Francisco L\'opez for developing the temperature control system), and to the IT people (specially Enrique De Guindos and Enrique de Juan). We also thank the referee, Dr. Francesco Pepe, for his thorough review of this manuscript and for the useful comments and suggestions that have improved its quality. J.L-B acknowledges financial support from the Marie Curie Actions of the European Commission (FP7-COFUND) and the Center for Astrobiology (CAB) for hosting me during the 4th year of the ESO Fellowship. J.L-B acknowledges support from the Spanish State Research Agency (AEI) through projects No.ESP2017-87676-C5-1-R and No. MDM-2017-0737 Unidad de Excelencia ÒMar\'ia de MaeztuÓ- Centro de Astrobiolog\'ia (INTA-CSIC). The active temperature controller has been funded thanks to the program so called 'Programa FEDER 2014-2020' of MINECO with reference ICST2017-07-CAHA-4 and CAHA-16-CE-3978 respectively. The program uses European Regional Development funds. The new grating has been funded thanks to the grant CAHA-15-CE-3902 of the program 'Acquisition of equipment for infrastructures' of the Ministerio de Econom\'ia y Competitividad (MINECO).




\bibliographystyle{mnras}
\bibliography{/Users/lillo_box/11_Mypapers/biblio2} 




\appendix

\section{Figures}
\newpage

\begin{figure*}
\centering
\includegraphics[width=1\textwidth]{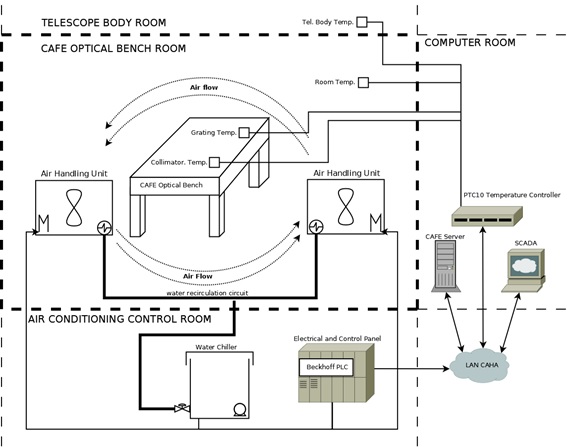}
\caption{Temperature controller for CAFE spectrograph. The figure shows the distribution of the sub-systems implemented.}
\label{fig:tempcontrol}
\end{figure*}

\begin{figure}
\centering
\includegraphics[width=0.5\textwidth]{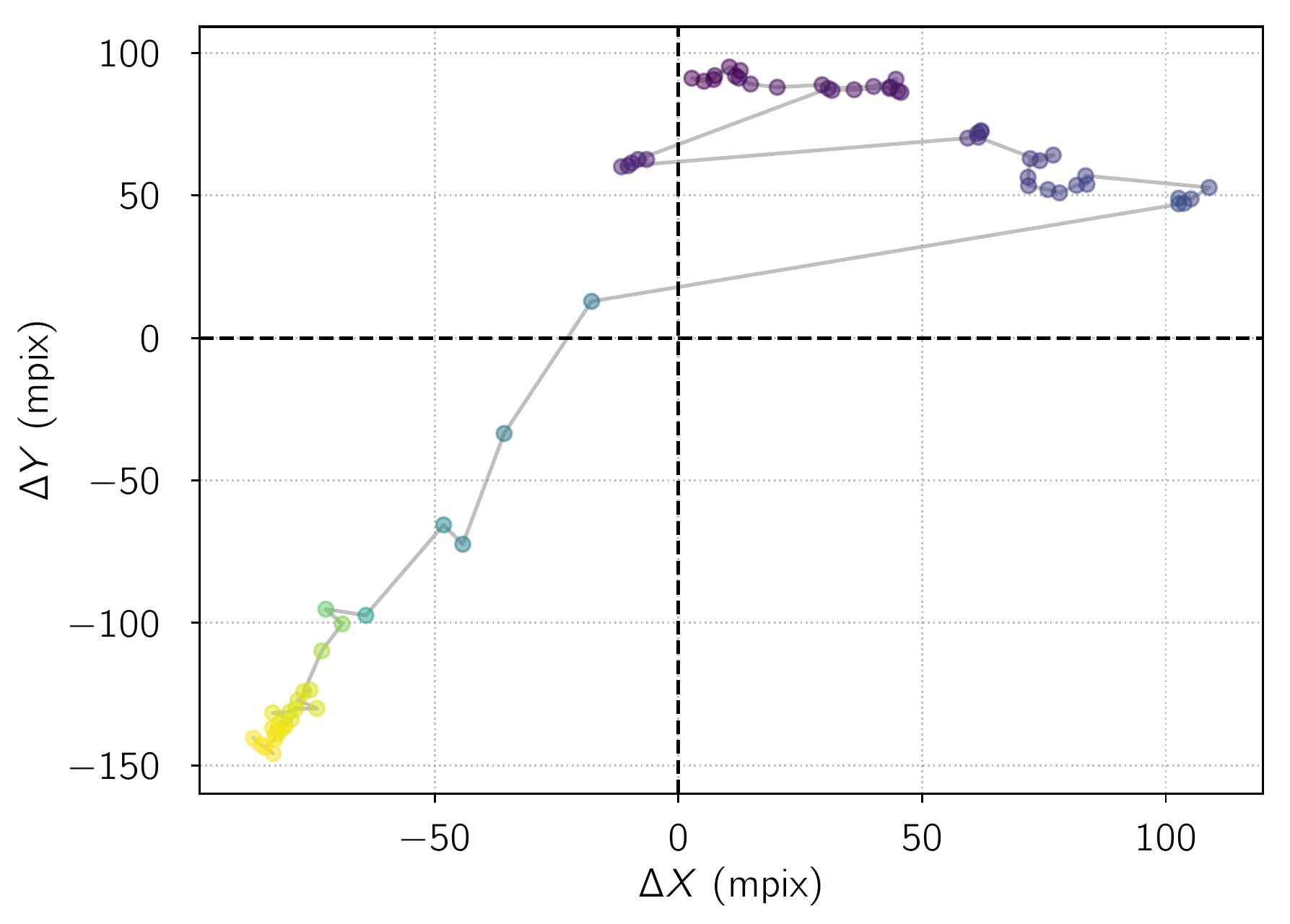}
\caption{Relative shifts in X and Y of the arc frames in one night respect to the reference frame during one of the commissioning nights in April 2019. The color-code of the data points correspond to the Julian date, indicating a clear night drift of the mean spot location.  }
\label{fig:arc_shifts}
\end{figure}

\begin{figure}
\centering
\includegraphics[width=0.5\textwidth]{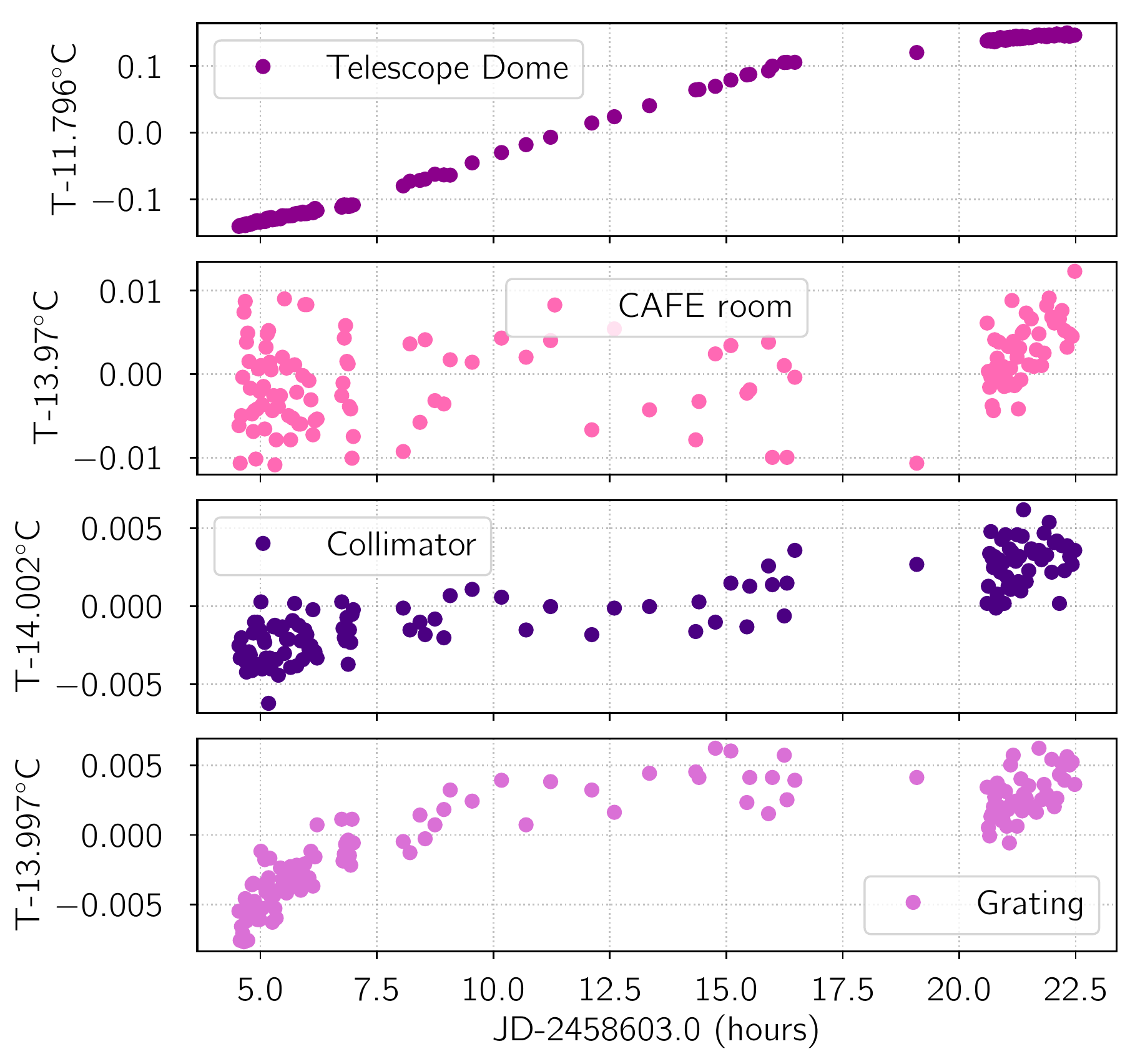}
\caption{Temperatures measured on the telescope dome and different locations and optical elements of the CAFE instrument as a function of time along one night.}
\label{fig:cafe_temps}
\end{figure}

\begin{figure*}
\centering
\includegraphics[width=1.0\textwidth]{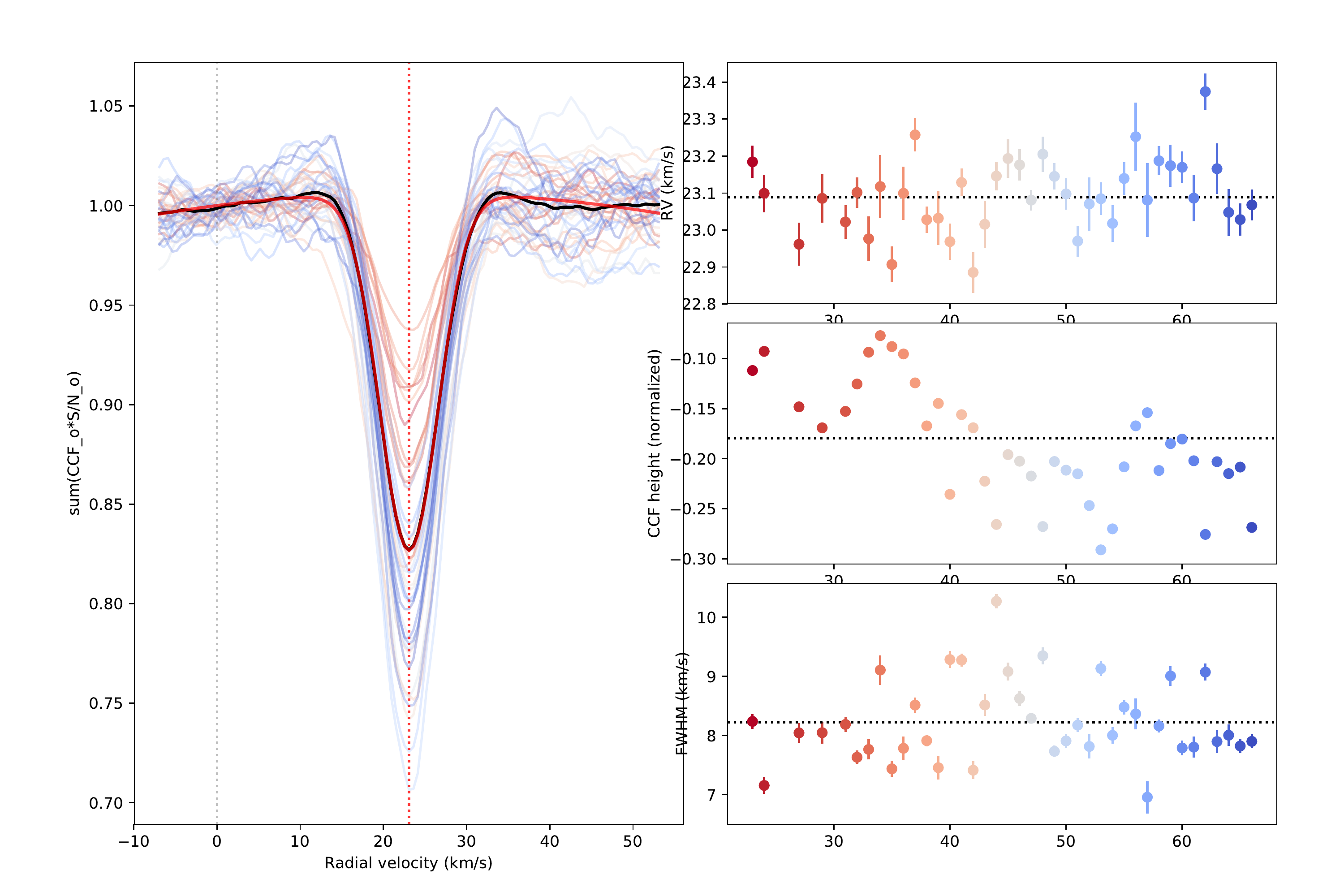}
\caption{Example of the pipeline plot data product corresponding to the CCF analysis of a RV standard. The left panel displays the CCF of the different orders (dark red to dark blue corresponding to orders from the red to the blue part of the spectrum), the final CCF in black thick line and the corresponding Gaussian fit in thick red line. The right panels show from top to bottom: RV, FWHM, and contrast for the CCF of the different orders.}
\label{fig:ccf}
\end{figure*}

\section{Tables}

\begin{table*}
\small
\setlength{\extrarowheight}{3pt}
\caption{Parameters of the setup file of \cafet and their default values.}              
\label{tab:defparams}      
\centering                                      
\begin{tabular}{l  l l}          
\hline\hline                        
Param. & Default	& Explanation	\\
\hline
root					& ~/path\_to\_data/			& Path of data folder \\
raw						& ~/path\_to\_data/RAW/		& Path of RAW data folder \\
redfolder				& ~/path\_to\_data/REDUCED/	& Path to REDUCED data folder to be created by pipeline \\
RefFrames				& ~/path\_to\_refFrames/		& Path to pipeline reference frames  \\
RefArc					& 'arc\_\_180718\_0031.fits'	& Reference arc filename \\
RefFlat					& 'flat\_\_180718\_0011.fits'	& Reference flat filename \\

bandwidth				& 1							& Temporal bandwidth to group calibration frames (in hours) \\
MinMembersBias 			& 4							& Minimum bias frames to create a Master bias \\
MinMembersFlat 			& 4							& Minimum flat frames to create a Master flat \\
MinMembersArcs 			& 9							& Minimum arc frames to create a Master arc \\
sigclip					& 3							& Sigma clipping for Master frames \\
							 
order\_aperture\_ampl 	& 5							& Amplitude of aperture in pixels for order extraction \\
order\_trace\_degree  	& 4							& Degree of polynomial for order tracing \\
							 
y0Nominal\_first 		& 115.433 					& Y-pixel of first order at central column [pix] \\
Nominal\_Nord 			& 79						& Number of orders to be exrtacted \\
ordID\_5500 			& 45						& Order corresponding to 5500\AA\  \\
order0 					& 62						& Real order corresponding to first extracted order \\

nx						& 3							& X-order of the polynomial for wavelength calibration \\
nm  					& 8							& Y-order of the polynomial for wavelength calibration \\
Selected\_MasterARC 	& 0 						& Default MasterArc to use for wavelength calibration \\

min\_extract\_col 		& 64						&  First column to be extracted \\
max\_extract\_col 		& 1984						&  Last column to be extracted \\
RO\_fl 					& 3.3						&  Readout noise \\
GA\_fl 					& 1.0						&  Gain of the CCD in e-/ADU\\

\hline                                             
\end{tabular}
\end{table*}

\begin{table*}
\small
\setlength{\extrarowheight}{3pt}
\caption{Spectral properties of CAFE and \cafet.}              
\label{tab:fitsformat}      
\centering                                      
\begin{tabular}{l  l l l l}          
\hline\hline                        
Ext.	& NAME	& TYPE	& SHAPE		& Comments \\
\hline
0	& PRIMARY     &   	PrimaryHDU  &    	-				& Header \\
1	& FLUX        &  	ImageHDU    &   	(2048, 84)  	& Extracted flux per order \\
2	& WAVELENGTH  &  	ImageHDU    &   	(2048, 84)  	& Wavelength matrix per order \\
3	& EFLUX       &  	ImageHDU    &   	(2048, 84)  	& Flux uncertainty per order \\
4	& FNORM       &  	ImageHDU    &   	(2048, 84)  	& Normalized flux per order \\
5	& EFNORM      &  	ImageHDU    &   	(2048, 84)  	& Uncertainty of normalized flux per order \\
6	& WMERGE1D    &  	ImageHDU    &   	(138432,)		& Wavelength array for the merged spectrum \\
7	& FMERGE1D    &  	ImageHDU    &   	(138432,)		& Flux array for the merged spectrum  \\
8	& CCF\_VEL    &  	ImageHDU    &   	(270,) 			& Cross-correlation velocities corresponding to CCF extension \\
9	& CCF         &  	ImageHDU    &   	(270,) 			& Cross-correlation function for velocities in CCF\_VEL extension \\
10	& ECCF        &  	ImageHDU    &   	(270,) 			& Uncertainty in CCF \\

\hline                                             
\end{tabular}
\end{table*}


\bsp	
\label{lastpage}
\end{document}